\documentclass{article}

\usepackage[final,nonatbib]{neurips_2020}

\usepackage[utf8]{inputenc} 
\usepackage[T1]{fontenc}    
\usepackage{hyperref}       
\usepackage{url}            
\usepackage{booktabs}       
\usepackage{amsfonts}       
\usepackage{nicefrac}       
\usepackage{microtype}      
\usepackage{float}
\usepackage{graphicx}
\usepackage{subfigure}

\usepackage{lipsum}

\newcommand\blfootnote[1]{%
  \begingroup
  \renewcommand\thefootnote{}\footnote{#1}%
  \addtocounter{footnote}{-1}%
  \endgroup
}

\title{dm2gal: Mapping Dark Matter to Galaxies with Neural Networks}

\author{%
  Noah Kasmanoff\\
  Center for Data Science\\
  New York University\\
  New York, NY 10011 \\
  \texttt{nsk367@nyu.edu} \\
   \And
   Francisco Villaescusa-Navarro\\
  Department of Astrophysical Sciences\\
  Princeton University\\
   Princeton NJ 08544 \\
  \texttt{fvillaescusa@princeton.edu} \\
  \And
   Jeremy Tinker\\
  Center for Cosmology and Particle Physics\\
  New York University\\
  New York, NY 10011 \\
  \texttt{jlt12@nyu.edu} \\
  \And
  Shirley Ho\\
  Center for Computational Astrophysics\\
  Flatiron Institute\\
  New York, NY 10010 \\
  \texttt{shirleyho@flatironinstitute.org} \\
}

\begin{document}

\maketitle

\begin{abstract}

Maps of cosmic structure produced by galaxy surveys are one of the key tools for answering fundamental questions about the Universe. Accurate theoretical predictions for these quantities are needed to maximize the scientific return of these programs. Simulating the Universe by including gravity and hydrodynamics is one of the most powerful techniques to accomplish this; unfortunately, these simulations are very expensive computationally. Alternatively, gravity-only simulations are cheaper, but do not predict the locations and properties of galaxies in the cosmic web. In this work, we use convolutional neural networks to \textit{paint} galaxy stellar masses on top of the dark matter field generated by gravity-only simulations. Stellar mass of galaxies are important for galaxy selection in surveys and thus an important quantity that needs to be predicted. Our model outperforms the state-of-the-art benchmark model and allows the generation of fast and accurate models of the observed galaxy distribution.\blfootnote{Code available at \hyperlink{https://github.com/nkasmanoff/dm2gal}{https://github.com/nkasmanoff/dm2gal}}

\end{abstract}

\section{Introduction}

Galaxies are not randomly distributed in the sky, but follow a particular pattern known as the \textit{cosmic web}. Galaxies concentrate in high-density regions composed of \textit{dark matter halos}, and galaxy clusters usually lie within these dark matter halos and they are connected via thin and long filaments. Those filaments are surrounded by very large low-density regions with almost no galaxies in them: cosmic voids.  Cosmologists use the cosmic web as a laboratory to learn about the fundamental laws and constituents of our Universe. The scientific community has invested billions of dollars in missions, both from ground and space, to survey the cosmic web as accurately as possible. In order to maximize the scientific return of those missions, accurate theoretical predictions are needed to extract the relevant information from observational data. Since these surveys observe galaxies and their properties such as stellar masses (the galaxy mass in stars), we need theoretical predictions for those quantities.

Cosmological hydrodynamic simulations are probably the best way to obtain these predictions; however, due to their large computational cost (millions of CPU hours), they only allow predictions of very small volumes. On the other hand, gravity-only simulations are much cheaper, but do not model galaxies nor their properties. In this work we try to bridge the gap between these two approaches using convolutional neural networks. Our purpose is to show that neural networks can learn to \textit{paint} galaxy properties on top of gravity-only simulations. This will speed up the process of creating predicted galaxy distributions used to analyze data from astronomical missions. In this work, we focus our attention in one of the most important galaxy properties, the \textit{stellar mass}, i.e. the mass in stars a galaxy contains.

The mapping we want to perform is
\begin{equation}
    M^h_*(\vec{x})=f(M^{g}_{\rm dm}(\vec{x}), M^{g}_{\rm dm}(\vec{y}))~,
\end{equation}
where $M_*^h(\vec{x})$ represents the stellar mass at position $\vec{x}$ according to the hydrodynamic simulation, $M^{g}_{\rm dm}(\vec{x})$ corresponds to the dark matter mass from the gravity-only simulation at position $\vec{x}$. We emphasize that the stellar mass of a galaxy will likely depend on its environment in a very complicated way. Although the underlying structure of the simulation pairs are the same, baryonic effects give rise to minor variations. That is the reason why we included the term $M^{g}_{\rm dm}(\vec{y})$ in the above equation, where $\vec{y}\ne\vec{x}$.Our purpose in this paper is to show that convolutional neural networks can approximate the function $f$.

Some of the studies that inspired this work are \cite{He_2019},\cite{zhang2019dark} and \cite{wadekar2020hinet}.

\section{Methods}

\subsection{Data}

We use data from the state-of-the-art magneto-hydrodynamic simulation TNG100-1 \cite{WeinbergerR16a,Pillepich2018}, and its gravity-only counterpart, TNG100-1-Dark, at present time. Those simulations contain, among other things, the position and mass of all particles in the simulations. Each simulation also contains a catalogue of dark matter halos with their properties (e.g. mass and position). We  construct the stellar mass and dark matter mass density fields from the particle positions and masses of the hydrodynamic and gravity-only simulations, respectively. Since galaxies are expected to reside in dark matter subhalos, we facilitate the training of the network by using also the mass-weighted subhalo field, that we construct from the gravity-only simulation. The fields span a volume of $(75~h^{-1}{\rm Mpc})^3$ ($1$ ${\rm Mpc}$ corresponds to $3.26$ million light-years) and they contain $2048^3$ voxels. 

One of the greatest challenges of working with this data is its sparsity: most of the voxels in this simulations do not contain galaxies (i.e. stellar mass is zero). We circumvent this problem by training the network only on regions centered on a subhalo with a stellar mass larger than $10^8~h^{-1}M_\odot$.

\subsection{Model}

Our network takes as input a two-channel 3D volume with $65^3$ voxels each: the dark matter and subhalos fields. The output of the model is the value of the stellar mass in the central voxel of the 3D fields. Our architecture consists of a series of blocks composed of convolutional, batch normalization, and ReLU activation layers that alternate between having a larger kernel size of $k\geq 5$ and stride 1, to a smaller kernel size, $k=3$, with stride 2. Both block types capture information on different scales, while efficiently down-sampling this large input to a single value. After six blocks, the latent representation is flattened into a vector that is passed through two fully connected layers which produces a predicted stellar mass. We will refer to this network as dm2gal, as its purpose is to map dark matter from gravity-only simulations to galaxies in hydrodynamic simulations.

\textbf{Weighted sampling.} The abundance of voxels with different stellar masses (the target) varies by many orders of magnitude. This poses a problem to our network, that learns to predict the stellar masses of the voxels with lowest stellar masses (the most abundant), but fails for the less frequent voxels with large stellar masses. We overcome this problem with a weighted sampling procedure. We first bin the voxels with stellar masses of the training data into $100$ bins logarithmically spaced between the minimum ($10^8~h^{-1}M_\odot$) and maximum ($10^{11.5}~h^{-1}M_\odot$) target values. We associate to each training sample a weight corresponding to the inverse the number count of values within its assigned bin. We also made use of data augmentation (3D random rotations) to increase our training data set.

\textbf{Training and validation.} From the $2048^3$ voxel fields, we reserve two cubes, one for validation and one for testing. The validation and testing cubes have $840^3$ ($30.76~h^{-1} \rm Mpc$) and $868^3$ ($31.78~h^{-1} \rm Mpc$) voxels, respectively. We save the model that best matched the cosmological statistics first on the validation cube, and then report performance on the testing cube. These regions were selected by requiring that they were representative enough, i.e. avoiding they contain big voids or very massive halos. The remaining voxels are used for training. 

We train our model by minimizing the mean square error $L_{\mathrm{MSE}} =  (\frac{1}{n})\sum_{i=1}^{n}(M_*^h(i) - M_*^{\mathrm{NN}}(i))^{2}$, where $M_*^h(i)$ and $M_*^{\mathrm{NN}}(i)$ are the stellar masses from the simulation and the prediction of the neural network, respectively. The sum runs over all samples in the training set. After trained to convergence,  we select for testing models that best match the stellar mass power spectrum of the validation region. Because the validation MSE value may correspond to good performance on only reconstructing low-mass galaxies, we avoid using it for indicating performance after training. 

\textbf{Hyper-parameter search.} We utilize PyTorch and PyTorch-Lightning \cite{falcon2019pytorch} to quickly train in parallel a broad range of hyper-parameter configurations, with learning rates between $10^{-5}$ to $10^{-1}$, weight decay between $0$ to $10^{-1}$, and capacity (number of channels in each layer). We employ a learning rate scheduler which decreases by a factor of $10$ for every 5 epochs in which the validation loss does not improve. Each model's best performing validation score was achieved within 24 hours of training on a single NVIDIA P100 GPU.

\subsection{Benchmark model}

We now describe the benchmark model we use to compare our results with. We refer to this method as HOD, from halo occupation distribution \cite{ScoSheHui01, Sel00, PeaSmi00, BerWei02}. The most important assumption of this model is that it considers that all galaxies reside within dark matter halos. The method works as follows. First, the dark matter halos from the hydrodynamic simulation are assigned to different bins according to their halo masses. Within each halo mass bin, galaxies are split into centrals and satellites, and their stellar mass distribution is calculated. Each halo mass bin will then have two stellar mass distributions: one for the centrals and one for the satellites. The HOD works as follows. We take a halo from the gravity-only simulation and its subhalos are split into central and satellites; the subhalo stellar masses are assigned by sampling the distribution obtained from the hydrodynamic simulation. We also correct for the effects of baryons on the halo mass function and number of satellites by multiplying the HOD prediction by the ratio of satellites, and overall halo mass between the simulations. We expect our HOD to perform better than the traditional one, where neither subhalo positions, nor halo mass corrections are considered.

\section{Results}

\begin{figure}
    \centering
    \includegraphics[width=14cm]{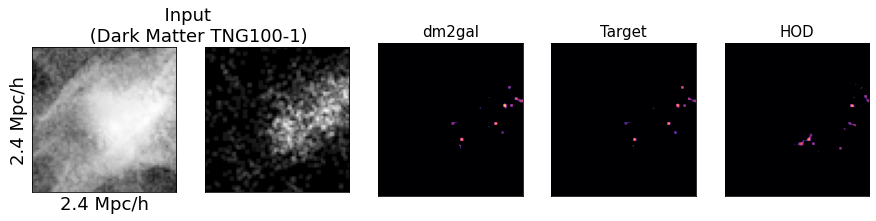}
    \subfigure{\includegraphics[width=4.6cm]{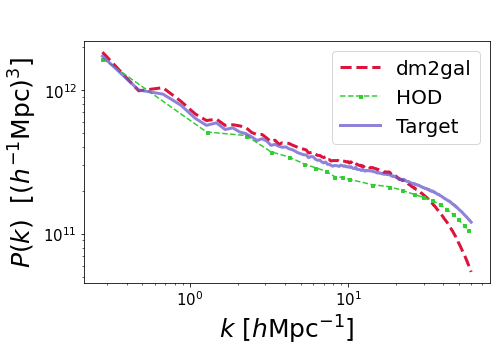}}
    \subfigure{\includegraphics[width=4.6cm]{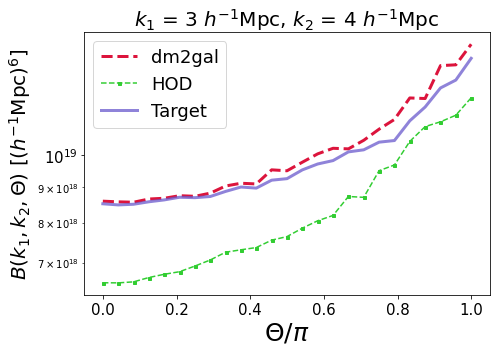}}
    \subfigure{\includegraphics[width=4.6cm]{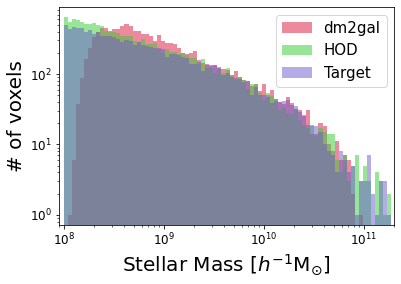}}
    \caption{The upper row shows the spatial distribution of dark matter and subhalos from the fast gravity-only simulations. Those fields are the inputs of the network, that outputs the stellar mass in the central voxel. By choosing different input regions the 3D stellar mass field can be predicted; we show it in the third panel. The fourth and fifth panels display the stellar mass from the expensive hydrodynamic simulation and the benchmark HOD model. The bottom panels compare different summary statistics (power spectrum-left, bispectrum-center, PDF-right) from the simulations (blue), dm2gal (red), and HOD (green). As can be seen, dm2gal outperforms the HOD in the clustering statistics (power spectrum and bispectrum) while yielding similar performance than the HOD for the relevant range of stellar masses.}
    \label{figure:plots}
\end{figure}

We now investigate the performance of our network on the test set, and compare the results against the HOD model. The first two panels of the upper row of Fig. \ref{figure:plots} show the spatial distribution of dark matter and subhalos from a $(2.4~h^{-1}{\rm Mpc})^3$ ($65^3$ voxels) region of the test set. These represent the input to the network, that outputs the value of the stellar mass in the central voxel. With inputs to the network at different spatial positions, the 3D stellar mass field can be predicted; we show it in the third panel. The stellar mass fields from the hydrodynamic simulation and the HOD model are shown in the fourth and fifth panels, respectively. From visual inspection, we find that dm2gal performs better than the HOD, and closely match the results of the hydrodynamic simulation. 

We now quantify the agreement between the predicted, HOD, and simulation stellar mass fields using three different summary statistics: 1) the power spectrum, 2) the bispectrum, and 3) the probability distribution function (PDF). 

\textbf{Power spectrum.} Given a 3D field, $\delta(\vec{x})$, we can compute its Fourier transform as $\delta(\vec{k})=\int e^{-i\vec{k}\cdot\vec{x}}\delta(\vec{x})d^3\vec{x}$ (using the discrete version for finite fields). The power spectrum can be computed as $P(k_i)=1/N_{k_i}\sum_{k\in[k,k+dk]} |\delta(\vec{k})|^2$, where $N_{k_i}$ is the number of independent modes in the internal $[k,k+dk]$. The power spectrum is one of the most important quantities in cosmology, as it describes the cosmic web on large, linear, scales. The first panel on the bottom row of Fig. \ref{figure:plots} shows the results. We find strong agreement on large scales (low values of $k$) between all fields for the power spectrum. This is expected for the HOD, but is a prediction for dm2gal. On smaller scales, dm2gal outperforms the HOD, with the exception of scales $k \geq 30~h{\rm Mpc}^{-1}$. We emphasize that these are extremely small scales (in cosmological terms), and most non-linear regime of physics. We believe that with more training on a higher resolution input, this fit will improve.

\textbf{Bispectrum.} The bispectrum is a higher-order statistic that contains non-Gaussian information from density fields \cite{Hahn_2019}. It is calculated as $B(k_1,k_2,\theta)=1/N_k\sum_{\vec{k}_1,\vec{k}_2|\vec{k}_1+\vec{k}_2+\vec{k}_3=\vec{0}}[\delta(\vec{k}_1)\delta(\vec{k}_2)\delta(\vec{k}_3)]$, where $N_k$ is the number of independent modes in the considered interval in $k_1$, $k_2$ and $\theta$, that is the angle between $\vec{k}_1$ and $\vec{k}_2$. We have taken a configuration with $k_1=3~h{\rm Mpc}^{-1}$ and $k_2=4~h{\rm Mpc}^{-1}$ and show the results of the bispectrum, as a function of $\theta$, in the middle panel of the bottom row of Fig. \ref{figure:plots}. In this case, we find that dm2gal outperforms the HOD for all angles. We have repeated the exercise for other triangle configurations, finding similar results.

\textbf{Probability distribution function.} Finally, we consider the probability distribution function, that we compute as the number of voxels with a certain stellar mass, as a function of the stellar mass value (for clarity, we do not normalize the distribution to have an area equal to 1 under it). This quantity contains additional information to the one embedded into the power spectrum and bispectrum \cite{Cora_2019}, and therefore, represents a different way to quantify the agreement between the different methods. We show the results in the bottom right panel of Fig. \ref{figure:plots}. We find that for stellar masses $M_* > 10^{8.5}~h^{-1}M_\odot$, both dm2gal and the HOD outputs a distribution very similar to the one from the hydrodynamic simulation. On the other hand, at the low mass end of the stellar mass PDF, the HOD outperforms dm2gal. We note that it is expected that the HOD model works very well for the PDF, as it is built to reproduce this statistic. We believe that with further training and tuning of the hyperparameters we can improve the results of the network in that regime. However, we emphasize that the low stellar mass regime is not very important for cosmological analysis, as astronomical surveys will have a hard time detecting such low mass objects.

\section{Conclusion}

We have shown, for the \textit{first} time, that convolutional neural networks can be used to \textit{paint} stellar masses into the dark matter field of computationally cheap gravity-only simulations. This method allows the production of stellar mass fields over large cosmological volumes. Generating these fields using hydrodynamic simulations will have a computational cost between 10x and 100x higher than with our method, that only requires running a gravity-only simulation. In terms of its performance, we have shown, that our model outperforms the traditional HOD method, while being more computationally efficient. 

This work has made use of simulations where the cosmological and astrophysics model is fixed. In the future, we plan to generalize our network to models with different cosmologies and astrophysics. We have also neglected any dependence on time that the mapping between dark matter to stellar mass may have. We plan to quantify the importance of that term by training the network using inputs at different times or by training using information from the merger trees. We also plan to extend the network to be able to predict other galactic properties such as metallicity, luminosity, and radius.

\section*{Acknowledgments}
We thank Jacky Yip, Carlos Fernandez-Granda, Gabriella Contardo, Yin Li, and Sigurd Naess for insightful discussions. This work was conducted using the computational resources at New York University and Princeton University

\section*{Broader Impact}

This work will benefit upcoming cosmological missions by speeding up the computational time needed to generate mock galaxy catalogues, needed to analyze the collected data. Before using this method for cosmological analyses, it is important to perform blind tests with simulations to corroborate that the network produce unbiased results for the required precision of the data. No ethical aspects are relevant for this work.






\bibliographystyle{unsrt}
\bibliography{main}

\end{document}